\documentstyle[12pt]{article}

\begin{document}
\renewcommand{\aa}{{\cal A}}

\begin{flushright}
NEIP-98-008\\
gr-qc/9808047 \\
August 1998
\end{flushright}

\begin{centering}
\bigskip
{\Large \bf Could we observe
the discreteness of \\
Quantum-Gravity length and area operators ?}\footnote{To appear in the 
proceedings of {\it 7-th International
Colloquium ``Quantum Groups and Integrable Systems"}, 
Prague, 18-20 June 1998.}

\bigskip
\medskip

{\bf Giovanni AMELINO-CAMELIA}\\
\bigskip
Institut de Physique, Universit\'e de Neuch\^atel,
     CH-2000 Neuch\^atel, Switzerland
\end{centering}
\vspace{1cm}
     
\begin{abstract}
Several proposals for Quantum Gravity involve
length and area operators with discrete eigenvalues.
I show that the analyses of some simple procedures
for the measurement of areas and lengths 
suggest that this discreteness characterizing the
formalism might not be observable.
I also discuss a possible relation with
the so-called $\kappa$ deformations of Poincar\'e symmetries.
\end{abstract}

\vspace{2cm}

The most fascinating problem that confronts the physics community 
is the one of reconciling Quantum Mechanics with Gravity.
The problem can be discussed using 
very familiar terms 
and we even have reasonably solid 
(although not conclusive \cite{Mtheory,mm})
arguments to infer 
that the Planck length, which is
nothing more than a combination of familiar scales,
should characterize the onset 
of observably large ``Quantum Gravity'' phenomena.
These familiar aspects of the problem have somehow
rendered even more frustrating
the lack of success of the decades of
efforts devoted to Quantum Gravity.
However, one should not forget that, although
the problem can be discussed in familiar terms,
the ``jump'' required in order to reach the Planck length 
($L_P \sim 10^{-33}cm$)
from presently-available experimental information (which can be roughly
characterized by the mass of the gauge bosons mediating the
weak interactions, {\it i.e.} of order $M_{W,Z}^{-1} \sim 10^{-16}cm$)
is gigantic even by comparison with previous major ``revolutions''
in our description of the physical world.
Trying to deduce the nature of Planck-length physics
from the experimental information we have presently available
is an extremely ambitious task, so much so that some of our colleagues
feel it would be futile.
While I shall not endorse this {\it a priori}
judgement of futility,
it is worth noticing that the comparison
of the relevant scales (sharpness of the probes versus
characteristic scale of the physics being studied)
indicates that one would be in better shape trying to discover and 
describe the weak interactions only using the information that can 
be gotten sitting in the stands of a stadium 
by watching a sporting event!

These observations should at least act as a warning not to trust
too much the intuition coming from presently-accessible
physical regimes in trying to
obtain a description of the Planck-length regime. In particular, the
successes of Quantum Mechanics provide no guarantee that all of its 
postulates should find use in the description of very short distances.
Within the Quantum-Gravity community it has been extensively debated 
whether ordinary Quantum Mechanics could be the appropriate
theoretical framework for Quantum Gravity. 
Some authors feel 
that we have convincing 
evidence (in the nature of the ``tension''
between Quantum Mechanics and Gravity) in support of
the fact that Quantum Mechanics should
be modified in order to accommodate Gravity
while other authors feel that the evidence
presently available is not solid enough to justify this conclusion. 
However, in light of the remarkable ``jump'' emphasized above
it seems that, even if we all agreed that at present we cannot
assume that Quantum Mechanics must be violated, we would still have 
to allow for the possibility that
such violations might emerge at some point down the road 
to the Planck scale. The problem becomes then the one of
getting some intuition for the nature of plausible candidates
for such violations, considering that presently attainable
experimental data can be of very little help.
It is in this respect that some recent analyses of gedanken experiments
can be most useful. A final test of the results of those analyses 
must wait for the correct Quantum Gravity or at least
some relevant experimental data, but those results already contribute
to the general development of Quantum-Gravity research by guiding us 
toward alternative scenarios which would otherwise not be considered.

In the present paper I discuss one of the ways in which Quantum-Gravity 
physics could violate some postulates of Quantum Mechanics.
Specifically, I shall be concerned with two properties of 
ordinary Quantum Mechanics: (1) that it allows a well-defined 
(although formal) limit in which
the devices used in measurements behave ``classically'', in the sense that 
their positions are completely under the control of the 
observer, and (2) that any given observable can be measured
with total accuracy (at the price of renouncing any information
on a conjugate observable) in the limit in which the 
devices composing the measuring apparatus behave classically.
In particular, the property (2)
would in principle allow to uncover 
any discreteness in the spectrum of a quantum operator.
This is quite important since many Quantum-Gravity scenarios
involve
length, area and volume operators which have discrete eigenvalues.
For example, 
an area operator~\cite{arsarea} with
discrete eigenvalues is one of the most intriguing aspects
of Canonical/Loop Quantum Gravity~\cite{canoni,loop,loopash},
which is a Quantum-Gravity approach that
(while being like all its
competitors only at the early stages of development and missing any 
experimental support)
has passed quite a few tests of formal and conceptual 
consistency.\footnote{Although in the context considered 
in Ref.~\cite{arsarea} areas
are not diffeomorphism-invariant,
the analysis reported by Rovelli in Ref.~\cite{rovarea}
suggests that a discrete spectrum should also
characterize areas specified
in a diffeomorphism-invariant
manner~\cite{mrsold,mrsrov} by matter fields.
In fact, in Ref.~\cite{rovarea} 
this discreteness was analyzed within
the model obtained by introducing matter fields
in the Husain-Kucha\v r Quantum-Gravity toy model~\cite{huskuch},
whose area operator is completely analogous to the one of
Canonical/Loop Quantum Gravity.}
Even more popular is the expectation that
the length operator should have discrete 
eigenvalues, at least in some Quantum-Gravity scenarios.

In this paper, I shall assume that indeed such discrete
length and area operators correctly describe physical lengths
and physical areas in the Quantum-Gravity
formalism, and investigate how this discreteness
that characterizes the formalism
would affect the outcome of experiments.
Some of the points made in the following are relevant for the
study of any diffeomorphism-invariant length and area operator
(whether or not the spectrum is discrete). Other aspects of 
the analysis apply only to diffeomorphism-invariant operators
with discrete spectrum, but still the details of the spectrum
are never important for the line of argument here proposed.
For simplicity, the reader can assume that the
length operator 
has eigenvalues ${\cal L}_n$ given by integer multiples
of the Planck length
\begin{eqnarray}
{\cal L}_n = n L_P ~,
\label{eigendist}
\end{eqnarray}
which is the type of quantization most commonly considered,
while the area operator 
has eigenvalues $\aa_n$ given by half-integer multiples
of the square of $L_P$:
\begin{eqnarray}
\aa_n = {n \over 2} L_P^2 ~,
\label{eigen}
\end{eqnarray}
which is the type of quantization found~\cite{rovarea}
in the Husain-Kucha\v r-Rovelli model.

Let us start by considering the
measurability of the distance $L$ between (the centers of mass of)
two bodies.
In Ref.~\cite{gacmpla} 
the measurement of the distance $L$ is discussed in terms
of the Wigner measurement procedure~\cite{wign,ng},
which relies on the exchange of a 
light signal between the two bodies.
The setup of the measuring apparatus schematically 
requires {\it attaching}
a light-gun, a clock, 
and a detector to one of the bodies
and {\it attaching} a mirror to the other body.
By measuring the time $T$ needed by the signal
for a two-way journey between the bodies one 
also
obtains a 
measurement of  $L$.
Within this setup it is easy to realize that 
$\delta L$ can vanish only if 
all devices used in the measurement behave classically.
One can consider for example the
contribution to $\delta L$ coming from 
the uncertainties that affect the relative motion of the clock
with respect to the center of mass of the system 
composed by the light-gun and the detector.
It is easy to show \cite{gacmpla,wign,ng} that
\begin{eqnarray}
\delta L \geq \sqrt{{ (M_c + M_{l + d}) \hbar T \over 
2 M_c \, M_{l + d} }}
~,
\label{dawign}
\end{eqnarray}
where $M_c$ is the mass of 
the clock and $M_{l+d}$ is the total mass of the system composed of
the light-gun and the detector.
Clearly, Eq.~(\ref{dawign}) implies that $\delta L = 0$ can
only be achieved in the ``classical-device limit,'' understood
as the limit of infinitely large $M_c$ and $M_{l+d}$.
This is consistent with the nature of the ordinary Quantum-Mechanics
framework, which relies on classical devices.
However, once gravitational interactions are taken into account
the classical-device limit is no longer available.
Large values of the masses $M_c$ and $M_{l+d}$ necessarily lead
to great distorsions of the geometry, and well before 
the $M_c , M_{l+d} \! \rightarrow \! \infty$
limit the Wigner measurement procedure can no longer be
completed. 
Since the classical 
limit $M_c , M_{l+d} \! \rightarrow \! \infty$ is not 
available,
from Eq.(\ref{dawign})
one concludes that 
in Quantum Gravity the uncertainty 
on the measurement of a length grows with 
the time $T$ required by the measurement procedure
(as it happens in 
presence of decoherence effects \cite{karo}).
In fact, Eq.(\ref{dawign}) can motivate \cite{gacmpla}
the expectation for a minimum uncertainty
for the measurement of a distance $L$ of the 
type
\begin{eqnarray}
minimum \left[ \delta L \right] \, \sim \, 
\sqrt{{ c T L_{QG}}} \,
\sim \, 
\sqrt{L \, L_{QG}}
~,
\label{gacup}
\end{eqnarray}
where $L_{QG}$ is a 
Quantum-Gravity length scale
that characterizes the mentioned limitations
due to the absence of classical devices,
and the relation on the right-hand side follows from
the fact that $T$ is naturally proportional~\cite{gacmpla,ng} 
to $L$.
Although $L_{QG}$ emerges in a way that does not appear
to be directly related to
the Planck length, it seems plausible \cite{gacmpla} 
that $L_{QG} \sim L_{P}$.

In the following I shall assume that indeed (\ref{gacup}) holds.

Let us now consider the measurement of areas.
I shall consider the measurement
procedure proposed by Rovelli in 
Ref.~\cite{rovarea}. There, for simplicity, the matter fields 
that specify the surface whose area is being measured
are taken to form a metal plate.
The area $\aa$ of this metal plate is 
measured using an electromagnetic device that keeps a second metal 
plate at a small distance $d$ and measures the capacity $C$ 
of the so formed capacitor.
Of course, measuring $d$ and $C$, and assuming 
that $d \ll \sqrt{\aa}$, one also measures $\aa$ as 
\begin{eqnarray}
\aa = C d ~,
\label{adc}
\end{eqnarray}
where I chose for simplicity units in which the 
relevant permittivity is 1.

According to Eq.~(\ref{adc}), in general 
the uncertainty in the measurement of the area $\aa$
receives contributions from the uncertainties in the determination
of $C$ and $d$.
Since I am aiming for a final result formulated as a measurability bound
({\it i.e.} a lower bound on the uncertainty),
it is legitimate to ignore the contribution coming from the
uncertainty in $C$ and focus 
on the contribution coming from the uncertainty in $d$
\begin{eqnarray}
\delta \aa \, \ge \, C \, \delta d \, = \, {\delta d \over d} \, \aa ~,
\label{deltadc}
\end{eqnarray}
where I also used again Eq.~(\ref{adc}) to eliminate $C$.

From the bound (\ref{gacup}) on the measurability of distances 
it follows that $\delta d /d \ge \sqrt{L_{QG}/d}$ and therefore
\begin{eqnarray}
\delta \aa \, \ge \, \sqrt{L_{QG}} \, {\aa \over \sqrt{d}}~.
\label{deltadc2}
\end{eqnarray}
This relation confronts us with a scenario similar to the one of
Eq.~(\ref{dawign}). 
It formally admits a 
limit ($d \rightarrow \infty$) in which
the area could be measured with complete accuracy, but this limit
cannot be reached within the constraints set by
the nature of the measurement procedure.
In fact, the relation (\ref{adc}), on which the measurement 
procedure is based,
only holds for $d \ll \sqrt{\aa}$, and in considering larger and
larger $d$ one quickly ends up loosing all information on $\aa$.
A rather safe lower bound is therefore 
obtained by imposing $d \le \sqrt{\aa}$
in Eq.~(\ref{deltadc2}), which gives
\begin{eqnarray}
\delta \aa \, \ge \, \sqrt{L_{QG}} \,\, \aa^{3/4} ~.
\label{deltadc3}
\end{eqnarray}

Bounds of the type
(\ref{gacup}) and (\ref{deltadc3}) 
would require a significant
shift in the
physical interpretation of 
quantization relations such as (\ref{eigendist}) and (\ref{eigen}).
In fact, assuming $L_{QG} \sim L_P$, Eq.~(\ref{gacup})
indicates that the measurement of a given length of order $n  L_P$
would be affected by an uncertainty of at least $\sim L_P (n)^{1/2}$,
{\it i.e.} (for every length with $n > 1$)
an uncertainty much larger than the $L_P$ quanta.
Similarly, assuming $L_{QG} \sim L_P$, Eq.~(\ref{deltadc3}) 
indicates that the measurement of a given area of order $n  L_P^2/2$
would be affected by an uncertainty of at least $\sim L_P^2 (n/2)^{3/4}$,
{\it i.e.} (for every area with $n > 1$)
an uncertainty much larger than the $L_P^2/2$ quanta.

Concerning the physical interpretation 
of Eqs.~(\ref{gacup}) and (\ref{deltadc3}) 
one is also naturally led to inquire about the type of symmetries 
that could result in such structures. Of course, 
it will be possible to rigorously address this question only
once a formalism supporting relations such as
(\ref{gacup}) and (\ref{deltadc3}) is found;
however, some consistency arguments~\cite{qgess98,kpoinpap}
appear to indicate that dimensionful deformations of 
Poincar\'e symmetries might be involved.
Interestingly, some of the predictions
of these deformations of Poincar\'e symmetries
could soon be tested \cite{grbgac}
experimentally by exploiting the recent dramatic 
developments in the phenomenology of gamma-ray bursts \cite{grbnews}.
While the interested reader should go to Refs.~\cite{qgess98,kpoinpap}
for a detailed discussion,
I shall here just sketch out one of the arguments
supporting the connection between dimensionful deformations of 
Poincar\'e symmetries and relations such as
(\ref{gacup}) and (\ref{deltadc3}).
I start by  observing that the
quantum $\kappa$-deformed Minkowski space~\cite{review,mr}
\begin{eqnarray}
~[x_j , x_k ] \!\!&=&\!\! 0
\label{commrelxx}\\
~[x_j , t ] \!\!&=&\!\! {x_j \over \kappa}
~, \label{commrelxt}
\end{eqnarray}
can be interpreted as implying that
the uncertainties on $x_j$ and $t$
satisfy
\begin{eqnarray}
\delta x_j \, \delta t \!\!&\ge&\!\! {x_j \over |\kappa|}
~. \label{newonegeneral}
\end{eqnarray}
This uncertainty relation 
has important implications 
for Wigner's measurement procedure.
In fact, interpreting (\ref{newonegeneral}) as a relation
between
the uncertainty $\delta t$ in the time when the light probe
sets off the clock at the end of its two-way journey
and 
the uncertainty $\delta x$ 
(along the axis of propagation of the light probe)
in the distance travelled by the probe by that same set-off time,
one obtains the relation~\cite{kpoinpap}
\begin{eqnarray}
\delta L \ge \delta x + c \, \delta t \sim \delta  x 
+ {2 c L \over |\kappa| \delta x}
~, \label{newadd}
\end{eqnarray}
which also takes into account 
that both $\delta x$ and $\delta t$ contribute to 
the total uncertainty in the measurement of $L$.
From (\ref{newadd}) one finds that
\begin{eqnarray}
\delta L \ge
\sqrt{ {2 c L \over |\kappa|}}
~,
\label{newbound}
\end{eqnarray}
which reproduces the relation (\ref{gacup}) 
upon appropriate association
of the scale $\kappa$ 
to the scale $L_{QG}$.

 


\end{document}